\documentclass[paperpaper, 10 pt, conference]{ieeeconf}  % Comment this line out if you need a4paper

\IEEEoverridecommandlockouts                              % This command is only needed if
\usepackage{graphicx}
\usepackage{booktabs}
\usepackage{multirow, multicol}
\usepackage{amsmath, amssymb}
\usepackage{algorithm, algorithmic}
\usepackage{textcomp}
\usepackage{booktabs,array}

\title{\LARGE \bf
Flocking and Collision Avoidance for a Dynamic Squad of Fixed-Wing UAVs Using Deep Reinforcement Learning}

\author{Chao Yan, Xiaojia Xiang, Chang Wang*, and Zhen Lan% <-this % stops a space
\thanks{This work was supported in part by the Science and Technology Innovation
2030-Key Project of New Generation Artificial Intelligence under Grant
2020AAA0108200, in part by the National Natural Science Foundation of
China under Grant 61825305, Grant 61803377, and Grant 61906203.}% <-this % stops a space
\thanks{College of Intelligence Science and Technology, National University of Defense Technology, Changsha 410073, China, \{yanchao17, xiangxiaojia, wangchang07, lanzhen19\}@nudt.edu.cn 
*Corresponding author}%
}

\begin{document}

\maketitle
\thispagestyle{empty}
\pagestyle{empty}
\hyphenpenalty=5000
\tolerance=1000
%%%%%%%%%%%%%%%%%%%%%%%%%%%%%%%%%%%%%%%%%%%%%%%%%%%%%%%%%%%%%%%%%%%%%%%%%%%%%%%%
\begin{abstract}
Developing the flocking behavior
for a dynamic squad of fixed-wing UAVs is still a challenge due to kinematic
complexity and environmental uncertainty. In this paper, we deal with the
decentralized flocking and collision avoidance problem through deep reinforcement
learning (DRL). Specifically, we formulate a decentralized DRL-based decision 
making framework from the perspective of every follower, where a collision
avoidance mechanism is integrated into the flocking controller. Then, we propose
a novel reinforcement learning algorithm PS-CACER for training a shared control
policy for all the followers. Besides, we design a plug-n-play embedding module
based on convolutional neural networks and the attention mechanism. As a result,
the variable-length system state can be encoded into a fixed-length embedding
vector, which makes the learned DRL policy independent with the number and the order of
followers. Finally, numerical simulation results demonstrate the effectiveness of
the proposed method, and the learned policies can be directly transferred to
semi-physical simulation without any parameter finetuning.
\end{abstract}

%%%%%%%%%%%%%%%%%%%%%%%%%%%%%%%%%%%%%%%%%%%%%%%%%%%%%%%%%%%%%%%%%%%%%%%%%%%%%%%%
\section{Introduction}
\label{sec1}
Due to the capability limitation of a single unmanned aerial vehicle (UAV),
multi-UAV collaboration has attracted increasing attention~\cite{ref1,ref2,ref_n1}.
One of the fundamental and challenging problems is
the flocking control of UAVs without collision~\cite{ref3}. Traditional
methods such as model predictive control~\cite{ref4} and consensus theory~\cite{ref5}
usually depend on precise physical models, which
are complex and difficult to obtain in practice.

As an alternative, reinforcement learning (RL)~\cite{ref6},~\cite{ref7} can be
used for the flocking control problem. For example, Hung et al. used the
Dyna-Q($\lambda{}$) algorithm~\cite{ref8} and the Q($\lambda{}$) algorithm~\cite{ref9}
to learn flocking control policies for fixed-wing UAVs.  Speck et al.~\cite{ref10} combined
the SARSA algorithm with object-focused learning to implement the formation control 
of UAV swarms. However, the above methods discretized the state and action spaces, 
which was inappropriate for controlling the UAVs in more realistic environments. 
Besides, the UAVs were assumed to fly at different fixed-altitudes to simplify the 
collision avoidance problem.

Deep reinforcement learning (DRL) has been proved effective for high-dimensional
and continuous control problems in robotics~\cite{ref11,ref12,ref13}. In our previous work, we proposed a DRL algorithm, i.e.,
continuous actor-critic with experience replay (CACER), for flocking with
fixed-wing UAVs in continuous state and action spaces~\cite{ref14,ref15}.
Following the previous work~\cite{ref8,ref9}, we also simplified the
collision avoidance problem by assuming that the UAVs maintained the same constant
speed but different altitudes. However, some real-world applications
such as surveying and mapping require the UAVs to flock at the same altitude.
This requirement makes flocking control more challenging, because the collision
avoidance problem must be considered. In contrast to the previous work
\cite{ref8},~\cite{ref9},~\cite{ref14}, this paper solves a more challenging flocking control
problem. Specifically, we not only allow the UAVs to change speed as needed, but
also design a collision avoidance mechanism and integrate it into the flocking
controller.

Besides, it is difficult to construct a leader-follower flocking control model for a dynamic squad of UAVs.
The main reason is that the length of the system state is relevant with the number of
followers, while the DRL-based control policies typically require a fixed-length
input for deep neural networks. Sui et al.~\cite{ref16} followed the method in~\cite{ref17}
where long short-term memory network (LSTM)~\cite{ref18} was applied to process
the states of other followers sequentially in the reverse order of their
distances to the decision-making follower, under the assumption that the nearest
neighbor had the biggest effect on the follower. However, this assumption is not
always true due to the influence of other factors such as speed and heading. Instead, we design a customized network module to deal with the collision avoidance problem for a dynamic squad of fixed-wing UAVs. 

The main contributions of this paper are as follows:

\begin{itemize}
	\item A decentralized DRL-based framework is developed to address the flocking control and
    collision avoidance problem for a dynamic squad of fixed-wing UAVs.
	\item A novel DRL algorithm is proposed for training the flocking controller, where a plug-n-play embedding module based on convolutional neural networks and the attention mechanism is designed to handle variable-length system state.
	\item The proposed method can be directly transferred from numerical simulation to
semi-physical simulation without any parameter finetuning.
\end{itemize}

The rest of this paper is organized as follows. Section~\ref{sec2} formulates the
flocking and collision avoidance problem, followed by the proposed PS-CACER
algorithm and the designed SEMP embedding module in Section~\ref{sec3}. Section~\ref{sec4}
discusses the simulation results. Finally, Section~\ref{sec5} concludes this paper.

\section{Problem Formulation}
\label{sec2}
In this section, we describe the flocking problem and the kinematics of fixed-wing UAV.
Then, we formulate this problem as a Markov decision process (MDP) in the RL framework.

\subsection{Problem Description}
\label{sec2.1}
\begin{figure}[thpb]
  \centering
  \includegraphics[width=2.5in]{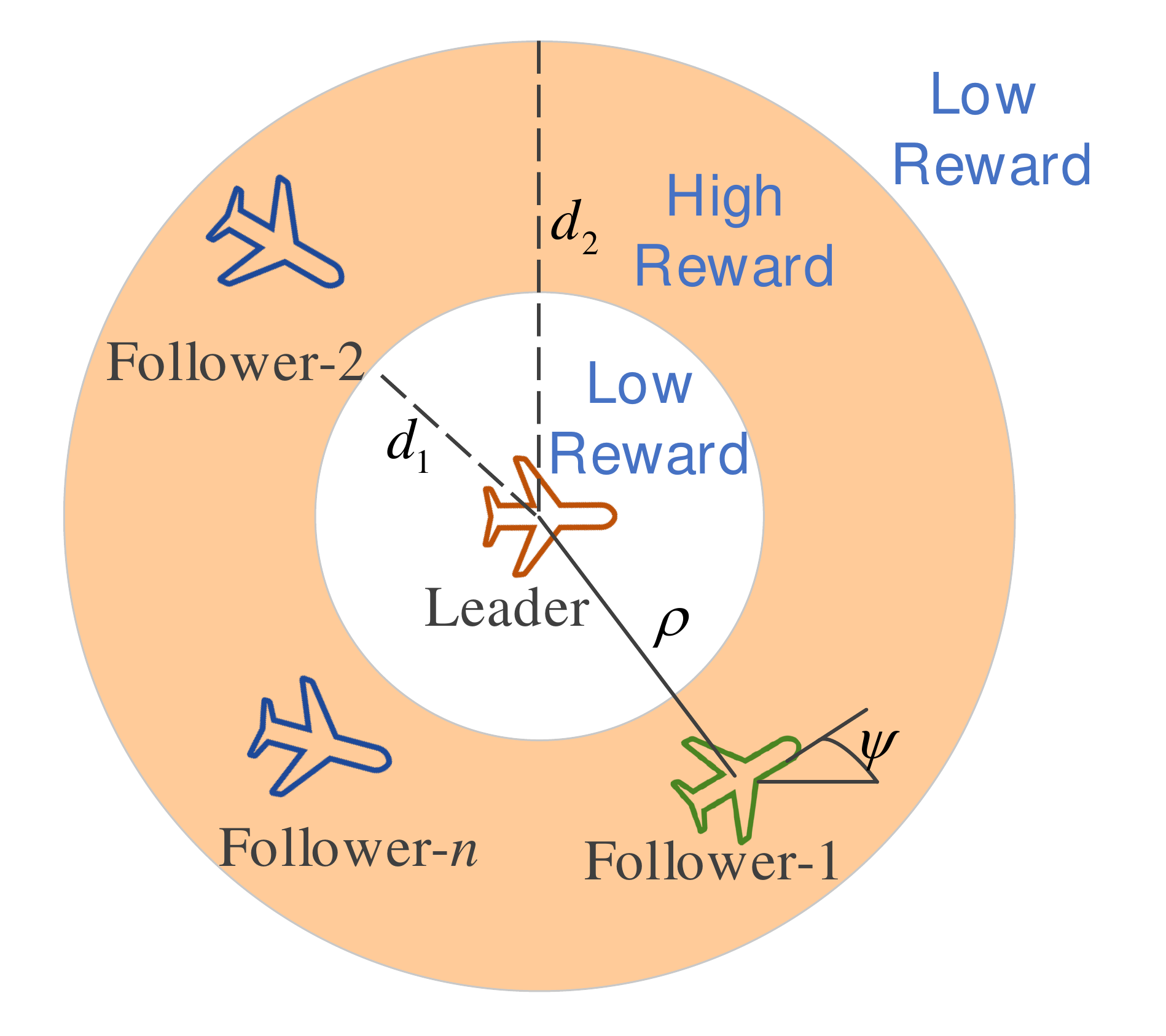}
  %\framebox{\parbox{3in}{fig1.jpg}}
  %\includegraphics[scale=1.5]{fig1.png}
  \caption{The top view of the relationship between the leader and followers. Note that the
number of followers, i.e., $n$, is a variable.}
  \label{figurelabe1}
\end{figure}

In our flocking scenario,  a unique leader is followed by a variable number of followers. These followers are homogeneous and fly at the same altitude~\cite{ref9, ref19}. The leader is remotely controlled by a human
operator via a ground control station. We assume that each UAV is able to obtain
the state of other UAVs through the inner communication channel~\cite{ref20}.
Each follower has to select its steering commands independently
to maintain a certain distance $\rho$ from the leader
( $d_1 < \rho < d_2$ ) while
avoiding collisions with other followers simultaneously, as shown in Figure~\ref{figurelabe1}.

\subsection{Kinematics of Fixed-Wing UAVs}

We use the same numerical kinematic model of
fixed-wing UAVs with stochastic disturbances as in~\cite{ref15} to generate
simulated samples for training the DRL-based control policy.
We rewrite the established kinematics as follows:
\begin{equation}
{\rm{\dot \xi }} = \frac{d}{{dt}}\left\{ {\begin{array}{*{20}{l}}
x\\
y\\
\psi \\
\phi \\
v
\end{array}} \right\} = \left\{ {\begin{array}{*{20}{c}}
{v\cos \psi  + {\eta _x}}\\
{v\sin \psi  + {\eta _y}}\\
{ - \left( {{\alpha _g}/v} \right)\tan \phi  + {\eta _\psi }}\\
{f_{\phi}\left( {\phi ,{\phi _d}} \right)}\\
{f_v\left( {v,{v_d}} \right)}
\end{array}} \right\}
,\label{eq:1}
%(1)
\end{equation}
where $\left( {x,\;y} \right)$ is the planar position; $\psi$ is the heading
angle; $\phi$ is the roll-angle; $v$ is the airspeed; ${\alpha _g}$ is
the gravitational acceleration; $\left( {{\eta _x},\;{\eta _y},\;{\eta _\psi }}
\right)$ are disturbance terms that follow normal distributions; the roll
dynamics $f_{\phi}\left( {\phi ,\;{\phi _d}} \right)$ and the airspeed dynamics
$f_v\left( {v,\;{v_d}} \right)$ describe the response relationship
between (i) the desired airspeed ${v_{d}}$ and the actual airspeed $v$,
and (ii) the roll-angle setpoint ${\phi _{d}}$ and the actual
roll-angle $\phi$, respectively.

\subsection{MDP of Flocking with Collision Avoidance}

\subsubsection{State Representation}

As mentioned above, there are one leader and $n$ followers. From the
perspective of an arbitrary follower, e.g., follower-$i$ ($i = 1, \
2, \ \ldots, \ n$), the flock of UAVs can be divided into three groups: the
follower-\textit{i} itself (termed as the \textit{ego-follower}), the leader, and
the other followers \{follower-$j$ $\vert$ $j = 1, \ 2, \ldots, \ n, \ j \not= i\}$. Accordingly, the system state is
composed of three parts: the ego-follower's state ${{\rm{\xi }}_e} = {\rm{\xi
}}_f^i$, the leader's state ${{\rm{\xi }}_l}$, and the other followers' states
${\rm{\xi }}_o = \{ {\rm{\xi}}_f^j$ $\vert$ $j = 1, \ 2, \ \ldots, \
n, \ j \not= i \}$, where ${\rm{\xi }}: =
(x,\;y,\;\psi ,\;\phi ,\;v)$ is a tuple that describes the state of a single
fixed-wing UAV according to its kinematics.

To reduce the redundancy of the representation of the system state, the \textit{ego-follower and leader joint state }${s^e}: = \left(
{s_1^e,\;s_2^e,\;s_3^e,\;s_4^e,\;s_5^e,\;s_6^e,\;s_7^e,\;s_8^e,\;s_9^e}
\right)$ is defined as:
\begin{equation}
\begin{aligned}\left[ {\begin{array}{*{20}{l}}
{s_1^e}\\
{s_2^e}
\end{array}} \right] &=\left[ {\begin{array}{*{20}{c}}
{\cos {\psi _l}}&{\sin {\psi _l}}\\
{ - \sin {\psi _l}}&{\cos {\psi _l}}
\end{array}} \right]\left[ {\begin{array}{*{20}{l}}
{{x_e} - {x_l}}\\
{{y_e} - {y_l}}
\end{array}} \right]\\
s_3^e &= {\psi _e} - {\psi _l} \\\vspace{0.5ex}
s_4^e &= {\phi _e} \\\vspace{0.5ex}
s_5^e &= {\phi _l} \\\vspace{0.5ex}
s_6^e &= \phi _{d}^l \\\vspace{0.5ex}
s_7^e &= {v_e} \\\vspace{0.5ex}
s_8^e &= {v_l} \\\vspace{0.5ex}
s_9^e &= v_{d}^l \end{aligned}
,\label{eq:2}
\end{equation}
where $(s_1^e,s_2^e)$ is the planar position of the ego-follower relative to the
leader; $s_3^e \in [-\pi, \pi)$ is the difference in the heading between the
leader and the follower. Besides, $\phi _{d}^l$
and $v_{d}^l$ are the roll-angle and velocity setpoints of the leader,
respectively.

The \textit{ego-follower and other followers joint state} $s_{}^o = \{ s_j^o: =
\left( {s_{1,j}^o,\;s_{2,j}^o,\;s_{3,j}^o,\;s_{4,j}^o,\;s_{5,j}^o}
\right)$$\vert{}$ $j = 1,\ 2,\ \ldots,\ n, \ j \not=
i\}$ is defined as:
\begin{equation}
\begin{aligned}\left[ {\begin{array}{*{20}{l}}
{s_{1,j}^o}\\\vspace{0.5ex}
{s_{2,j}^o}
\end{array}} \right] &=\left[ {\begin{array}{*{20}{c}}
{\cos {\psi _e}}&{\sin {\psi _e}}\\\vspace{0.5ex}
{ - \sin {\psi _e}}&{\cos {\psi _e}}
\end{array}} \right]\left[ {\begin{array}{*{20}{l}}
{{x_f^j} - {x_e}}\\\vspace{0.5ex}
{{y_f^j} - {y_e}}
\end{array}} \right]\\
s_{3,j}^o &= {\psi _f^j} - {\psi _e}\\\vspace{0.5ex}
s_{4,j}^o &= {\phi _f^j}\\\vspace{0.5ex}
s_{5,j}^o &= v_{f}^j \\ \end{aligned}
,\label{eq:3}
\end{equation}
where  $\left( {s_{1,j}^o,s_{2,j}^o} \right)$ represents the planar position of
another follower-$j$, relative to the ego-follower; $s_{3,j}^o \in [-\pi, \pi)$ denotes
the difference in the heading between the ego-follower and follower-$j$.

Consequently, concatenating the two joint states above, we construct the state
representation of the system $s: = \left( {s_{}^e,\;s_{}^o} \right)$. We note
that its dimensionality depends on the number of followers.

\subsubsection{Action Space}
\label{sec2.1.1}
The followers are maneuvered by executing the roll and velocity actions.
In this paper, we define the action  $a: = \left( {a_r^{},\;a_v^{}} \right)$ in
continuous spaces, where the roll action  ${a_r} \in [-\frac{\pi}{18}, \frac{\pi}{18}]$ and the velocity action ${a_v} \in \left[ { - 1,\; + 1}
\right]$. The current roll-angle and airspeed of a follower are denoted by $\phi$ and $v$, respectively. The next roll-angle setpoint  ${\phi_{d}}$ can be calculated by:
\begin{equation}
\phi_{d}=\left\{\begin{array}{ll}\vspace{0.5ex}
r_{\mathrm{bd}} & \mathrm { if } \ \phi+a_{r}>r_{\mathrm{bd}} \\\vspace{0.5ex}
-r_{\mathrm{bd}} & \mathrm { if } \ \phi+a_{r}<-r_{\mathrm{bd}} \\\vspace{0.5ex}
\phi+a_{r} & \mathrm { otherwise }
\end{array}\right.
,\label{eq:4}
\end{equation}
where $[ - {r_{{\rm{bd}}}},\;{r_{{\rm{bd}}}}]$ is the allowed range of the
roll-angle setpoint.\\
Similarly, the next velocity setpoint ${v_{d}}$ is defined by:
\begin{equation}
v_{d}=\left\{\begin{array}{ll}\vspace{0.5ex}
{v_{\max}} &  \mathrm {if}\ v+a_{v}>{v_{\max}}  \\\vspace{0.5ex}
{v_{\min}}  &  \mathrm {if}\ v+a_{v}<{v_{\min}}  \\\vspace{0.5ex}
v+a_{v} &  \mathrm { otherwise }
\end{array}\right.
,\label{eq:5}
\end{equation}
where the ${v_{\max }}$ and ${v_{\min }}$ denote the maximum velocity and the
minimum velocity of the follower, respectively.

\subsubsection{Reward Function}

In this paper, the followers aim to avoid collision between each other while
flocking with the leader. Therefore, the reward function consists of two pieces:
the \textit{flocking reward }and the \textit{collision penalty}. Specifically, in
order to facilitate the agent (i.e., the ego-follower) to maintain a suitable
distance from the leader, the \textit{flocking reward} ${r_l}$ is designed as follows~\cite{ref9}:
\begin{equation}
r_l =  - \max \left\{ {{d_e},\;\frac{{{d_1}\left| {s_3^e} \right|}}{{\pi
\left( {1 + \omega {d_e}} \right)}}} \right\}
,\label{eq:6}
\end{equation}
where ${d_e} = \max \left\{ {m\left( {{d_1} - \rho } \right),0,\rho  - {d_2}} \right\}$. $d_{1}$ and $d_{2}$ are the inner and outer radius of the desired annulus shown in Figure~\ref{figurelabe1}, respectively. $\rho  = \sqrt {{{\left| {s_1^e} \right|}^2} + {{\left| {s_2^e} \right|}^2}}$ denotes the distance
between the leader and the agent. Both $\omega$ and $m$ are tuning
parameters. 

Besides, the \textit{collision penalty} $r_c^j$ is defined to prevent the agent
from colliding with the follower-$j$, as follows:
\begin{equation}
r_c^j = - \max \left\{ {m\left( {{d_1} - \rho _j^o} \right),0} \right\}
,\label{eq:7}
\end{equation}
where $\rho _j^o = \sqrt {{{\left| {s_{1,j}^o} \right|}^2} + {{\left| {s_{2,j}^o} \right|}^2}}$ is the distance from the agent to follower-$j$.

Lastly, the final reward function $r$ is specified as:
\begin{equation}
r = {r_l} + \sum\nolimits_j {r_c^j}
.\label{eq:8}
\end{equation}

\section{Approach}
\label{sec3}

\subsection{PS-CACER}

In our previous work~\cite{ref14},~\cite{ref15}, we proposed an algorithm
called continuous actor-critic with experience replay (CACER). In contrast to
other actor-critic algorithms, e.g., DDPG~\cite{ref22}, one distinctive feature of
CACER is in its positive-temporal difference (TD) update scheme~\cite{ref23}
for training the policy (actor). In other words, CACER updates its
policy only when the TD-error is positive~\cite{ref24}. In this paper, we
propose a novel algorithm PS-CACER by extending CACER to a multi-agent scenario
(see Algorithm~\ref{alg1}). Specifically, PS-CACER has a plug-n-play embedding module (see Section~\ref{sec3.2}) based on convolutional neural networks and the attention mechanism to handle variable-length system state, while CACER only deals with fixed-length system state.

\begin{algorithm}[t!]
\caption{PS-CACER}
\renewcommand{\algorithmicrequire}{\textbf{Input:}}
\renewcommand{\algorithmicensure}{\textbf{Output:}}
\label{alg1}
\begin{algorithmic}[1]
{\small\REQUIRE $N_{s} \ -$ maximum time steps; $N_{b} \ -$ training batch size;\\
\hspace{10pt} $M \ -$ desired number of training episodes}
\STATE {\small Empty replay memory $D$ with capacity $N$}
\STATE  {\small Initialize actor $Act_{\theta ^A} (s)$ and critic $V_{\theta ^V}(s)$ randomly}
{\small \FOR {$episode = 1$ to $M$}
    \STATE Initialize the number of followers $n$ randomly
    \STATE Represent initial state $s_i$ for each follower $i$
    \FOR {$t = 1$ to $N_{s}$}
        \FOR {follower $i = 1$ to $n$}
           \STATE Select action ${a_i} \leftarrow Act_{\theta^A}\left(s_i\right) + \mathcal N$ with the current policy and Gaussian exploration  noise
           \STATE Execute the selected action ${a_i}$
        \ENDFOR
        \FOR {follower $i = 1$ to $n$}
           \STATE Represent new system state ${s'_i}$
           \STATE Calculate immediate reward ${r_i}$
           \STATE Store tuple $({s_i},{a_i},{r_i},s'_i)$ in $D$
           \STATE $s_{i} \leftarrow s_{i}^{\prime}$
        \ENDFOR
        \STATE Sample a minibatch of $N_{b}$ tuples $({s_k},{a_k},{r_k},s'_k)$ from $D$
        \STATE  Empty temporal buffer $D' = \varnothing$
        \FOR {tuple $k = 1$ to $N_{b}$}
           \STATE Calculate TD-error: ${\delta _k} = {r_k} + \gamma  \cdot V_{\theta^V}\left({s'}_k \right) -
           V_{\theta^V}\left(s_k\right)$
           \STATE Store tuple $k$ to $D'$ if ${\delta _k} > 0$
        \ENDFOR
        \STATE Optimize actor by minimizing the loss: \\$\frac{1}{{\left\| {D'} \right\|}}{\sum\nolimits_{k'} {\left\| {{a_{k'}} - Act_{\theta^A}\left( {s_{k'}} \right)} \right\|}^2}$
        \STATE  Optimize critic by minimizing the loss of  $\frac{1}{{{N_b}}}{\sum\nolimits_k {\left\| {{\delta _k}} \right\|} ^2}$
    \ENDFOR
\ENDFOR}
\end{algorithmic}
\end{algorithm}

As mentioned in Section~\ref{sec2.1}, the followers are homogeneous. Thus, we adopt 
the parameter sharing (PS) approach~\cite{ref25} and allow all the followers to share 
the parameters of a single policy. This shared policy can be optimized more efficiently 
with the data of experiences collected by all the followers. We note that the
PS-CACER algorithm follows the \textit{centralized-learning and
decentralized-execution} fashion~\cite{MADDPG}. In other words, during the
learning phase, experiences obtained by all the followers simultaneously are
stored into a shared experience replay memory, and the shared policy in terms of
deep neural networks are trained with the stored experiences centrally
(centralized-learning). Nevertheless, during the execution phase, each follower
independently selects and executes its action by following the shared policy
based on its perceived state (decentralized-execution).

\subsection{Network Architecture}
\label{sec3.2}

Inspired by
\cite{ref26} and~\cite{ref27}, we design a customized network module based on convolutional neural networks and attention mechanisms, which enables our algorithm to adapt to the changes in the number of followers.

\begin{figure}[thpb]
  \centering
  \includegraphics[width=2.9in]{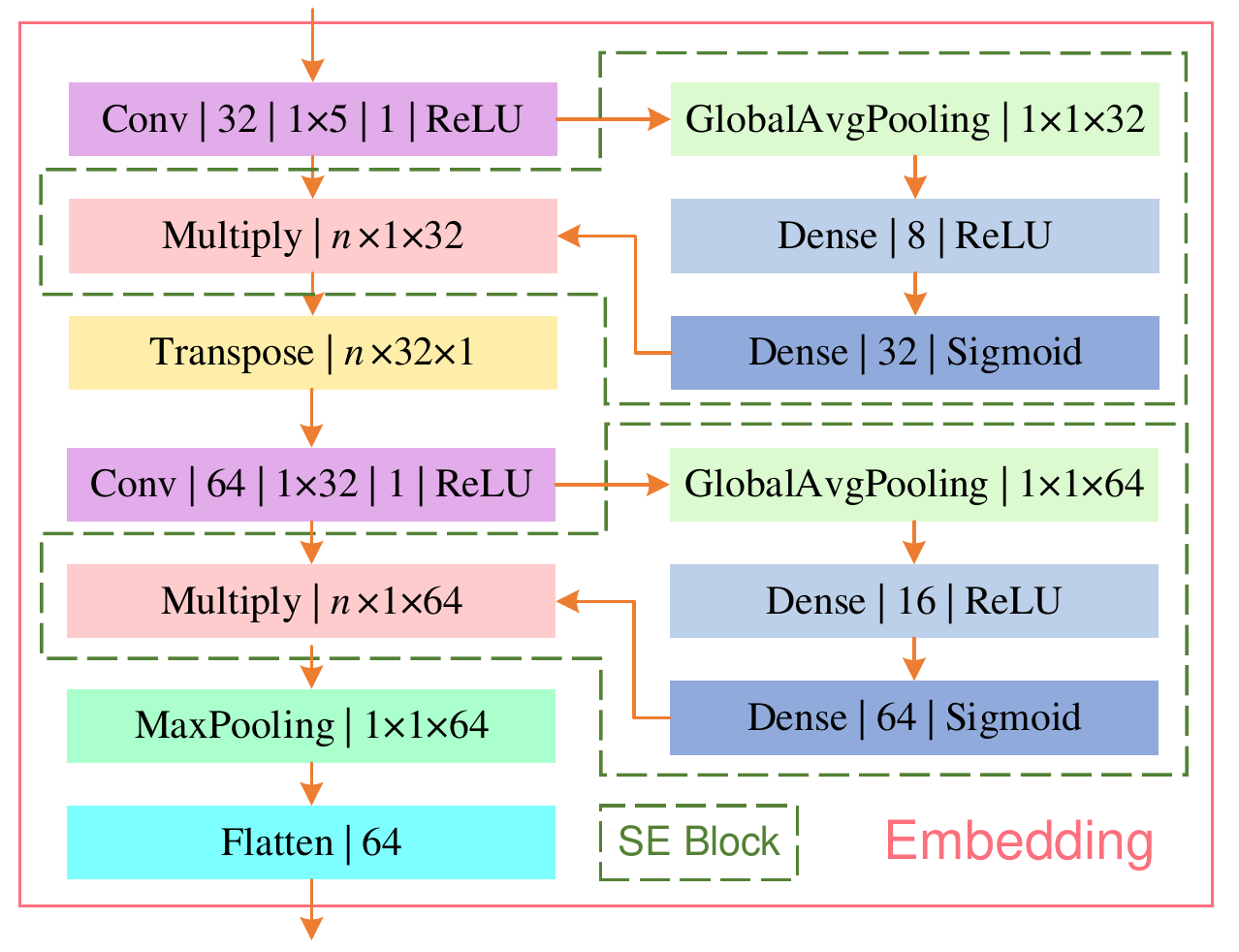}
      %\framebox{\parbox{3in}{fig1.jpg}}
      %\includegraphics[scale=1.5]{fig2.png}
  \caption{Embedding module. Each convolutional layer (Conv) is featured by the number of filters, filter size, stride, and activation mode. Other layers are represented by their type/name and output size.}
  \label{figurelabe2}
\end{figure}

As illustrated in Figure~\ref{figurelabe2}, the Embedding module is composed of two
convolutional layers (Conv), two \textit{Squeeze-and-Excitation} (SE) blocks
\cite{ref27}, a transpose layer (Transpose), a max-pooling layer (MaxPooling), and a
flatten layer (Flatten). First, the variable-length input (i.e., the
\textit{ego-follower and other followers joint state }${s^o}$) is passed by two
convolutional layers (i.e., Conv1 and Conv2) successively. The filter size of
Conv1 is equal to the length of $s_j^o$ , while the filter size of Conv2 is equal
to the number of filters of Conv1. With this specialized structure, each feature
map extracted by convolutional layers only depends on one follower (one of the
other followers). Besides, benefitting from the properties of CNNs, i.e., weight
sharing and shift invariance~\cite{ref27}, the sequencing operations according to the distances between
the leader and the followers as in~\cite{ref16,ref17} are no longer required. In
other words, the output are invariant to the indexing of the followers.

\begin{figure}[thpb]
  \centering
  \includegraphics[width=3.1in]{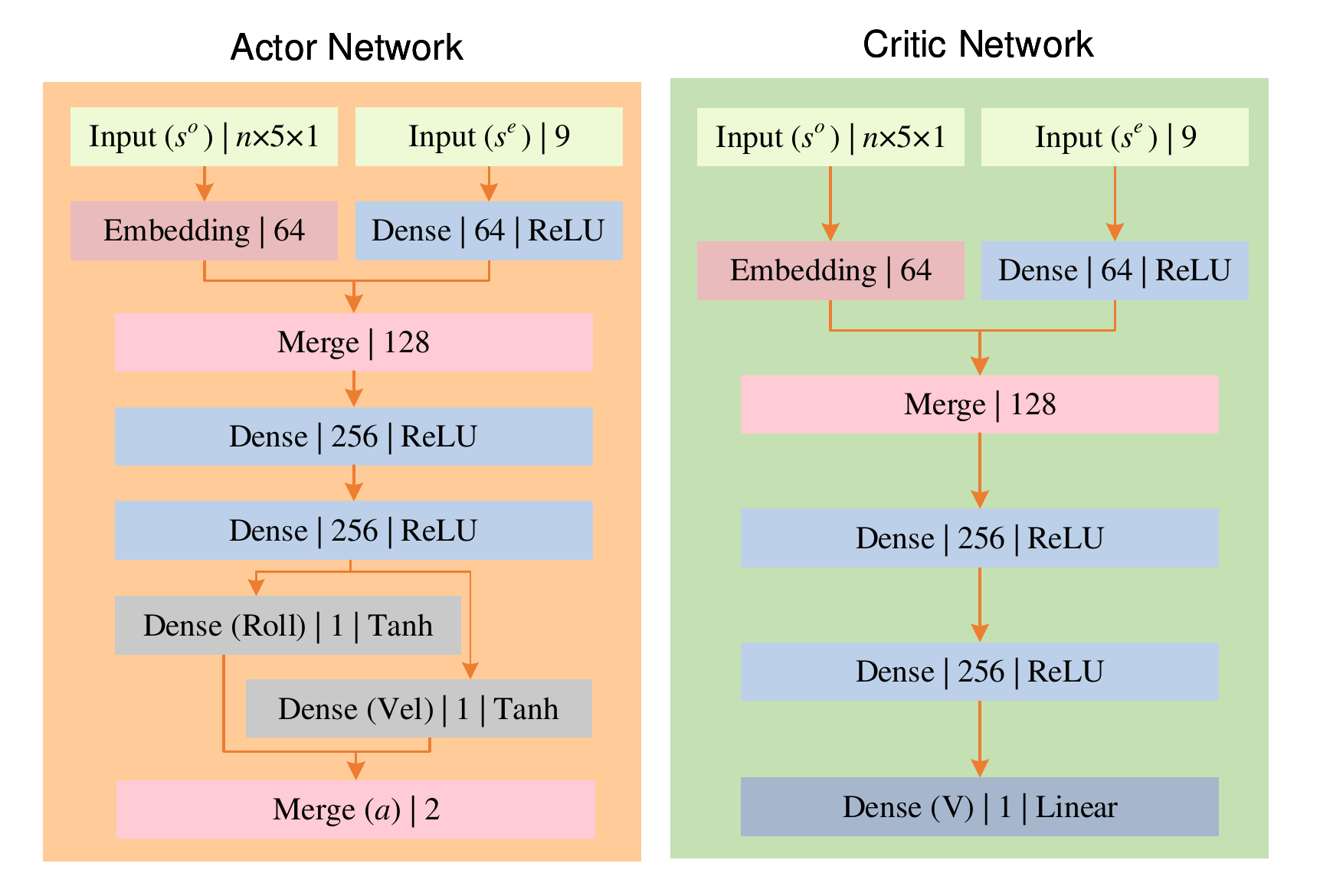}
      %\framebox{\parbox{3in}{fig1.jpg}}
      %\includegraphics[scale=1.5]{fig2.png}
  \caption{Network architecture. Each fully-connected layer (Dense) is featured by its type, number of neurons, and activation mode. Other layers are represented by their type/name and output size.}
  \label{figurelabe3}
\end{figure}

After each convolutional layer, a SE block is added to improve the capacity for
feature extraction of networks. The SE block is a channel attention mechanism. Each SE block uses a global
average pooling layer (GlobalAvgPooling) in the \textit{squeeze} phase and two
fully-connected layers (Dense) with different activation functions in
\textit{excitation} phase, followed by a channel-wise scaling operation
(Multiply). By explicitly modelling the relationship between channels, this
architectural unit can enhance its selectivety to the more informative features
while suppressing the less useful ones, boosting the representational power of
the network~\cite{ref27}. After the second Multiply layer, a max-pooling layer
is added to create a fixed-size output, independent of the input order and
length. Lastly, the output of the max-pooling layer is flattened by a flatten
layer to create a fixed-length embedding vector. In this way, the
variable-length input can be embedded into a
fixed-length embedding vector.

We term our scheme of the Embedding module as \textit{SE-MaxPooling} (SEMP). The
designed SEMP embedding module can handle inputs with arbitrary length and its
output is independent of the input order. We emphasize that this module
is plug-n-play, which means the SEMP embedding module can be easily integrated with
other network architectures, as well as any reinforcement learning algorithm.

The entire architecture of the actor network and the critic network is depicted
in Figure~\ref{figurelabe3}. We process the \textit{ego-follower and leader joint state ${s^e}$}
and the \textit{ego-follower and other followers joint state ${s^o}$}
separately~\cite{ref16}. Specifically, a dense layer with ReLU
activation function is used to extract the features of ${s^e}$, and the Embedding
module is used to encode the ${s^o}$ into a fixed-length vector. After that, the
two outputs above are merged and then subsequently fed into 3 dense layers with
different activation functions. Note that both the actor and the critic use the
same network architecture up to the last output layer. The output layer of the
critic uses a linear activation function, while the output payer of the actor
uses a hyperbolic tangent (tanh) activation function.

\section{Simulation Results}
\label{sec4}

In this section, we evaluate the proposed PS-CACER algorithm with the SEMP
scheme in both numerical simulation and semi-physical simulation.

\subsection{Simulation Setup}

In this paper, the PS-CACER algorithm was trained with a total of 30000 episodes ($M = $ 30000),
in which each episode had a maximal number of 60 time steps ($N_{s} = $ 60). All network parameters
were updated with 64 batch size ($N_{b} = $ 64), Adam optimizer with 0.001 (for actor) and 0.0001
(for critic) learning rates. The exploration parameter was
annealed exponential from 0.5 to 0.05 over a period of 2000 episodes, and then
fixed at 0.05 thereafter. The empirical values of essential
parameters are listed in Table~\ref{tab1} ~\cite{ref14, ref15}.

\begin{table}[htp]
	\centering
	\caption{Parameter Settings}
    \renewcommand{\arraystretch}{1.1}
    \label{tab1}
    \setlength{\tabcolsep}{6mm}{
	\begin{tabular}{|c|c||c|c|}
		\hline
		\textbf{Name} & \textbf{Value} & \textbf{Name}& \textbf{Value} \\
		\hline
		\textit{d}$_{1}$ & 40 & \textit{d}$_{2}$ & 65 \\
		\hline
		\textit{$\omega $} & 0.05 & \textit{m} & 2 \\
        \hline
		\textit{${\alpha _g}$} & 9.8 & \textit{${r_{{\rm{bd}}}}$} & $\frac{\pi}{6}$ \\
		\hline
        \textit{${v_{\max }}$} & 18 & \textit{${v_{\min }}$} & 12 \\
		\hline
        $N$ & 100000 & \textit{$\gamma $} & 0.95 \\
		\hline
	\end{tabular}}
\end{table}

\subsection{Numerical Simulation}
In order to verify the effectiveness of the proposed SEMP scheme, three existing
state-of-the-art approaches, LSTM~\cite{ref16}, SA (social attentive pooling)
\cite{ref29}, and CNNMP (convolutional neural networks with max pooling)\cite{ref26},
were selected as the benchmarks. To make a fair comparison, we kept
their network architectures the same except for the embedding module. Besides,
the above solutions also used the same learning  algorithm, i.e., the PS-CACER
algorithm, to update their network parameters. We note that, i) the number of LSTM cells was set to 64, in accordance with the
length of the fixed-length vector encoded by SEMP. ii) the network and its
parameters used by SA were the same as~\cite{ref29}. iii) The network structure
of CNNMP was the same as SEMP, except that SEMP merged a channel attention
mechanism by adding two SE blocks.

In the training phase, we evaluated the performance of the proposed algorithm
using the \textit{average reward} $G_{{\rm{Avg}}}$ obtained by one agent within
a certain number of $n$$_{e}$ episodes:
\begin{equation}
{G_{{\rm{Avg}}}} = \frac{1}{{n{\kern 1pt} {N_e}{N_s}}}\sum\limits_{i = 1}^n
{\sum\limits_{p = 1}^{{N_e}} {\sum\limits_{t = 1}^{{N_s}} {{}^ir_t^p} } }
,\label{eq:9}
\end{equation}
where ${}^ir_t^p$ is the immediate reward obtained by agent $i$ at the time
step $t$ of the episode $p$ according to (8). Without loss of generality, we set $N_{e} = $ 100. At each episode, the number of followers
$n$ was randomly selected from 3 to 10. At each time step, the initial state of the leader and followers were initialized
randomly, and the steer commands of the leader were generated randomly from the  action spaces defined in Section~\ref{sec2.1.1}.

\begin{figure}[b!]
  \centering
  \includegraphics[width=3.in]{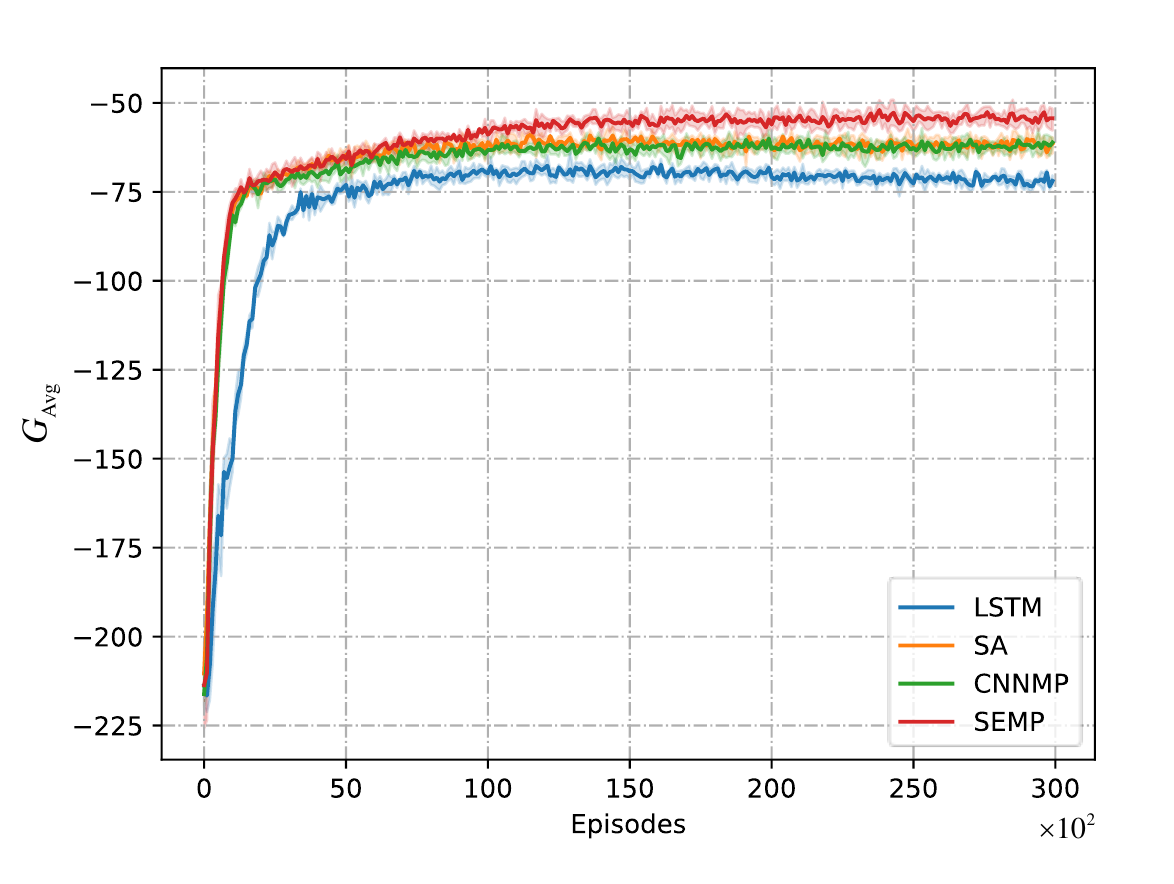}
      %\framebox{\parbox{3in}{fig1.jpg}}
      %\includegraphics[scale=1.5]{fig2.png}
  \caption{Learning curves of LSTM, SA, CNNMP, and SEMP.}
  \label{figurelabe4}
\end{figure}

\begin{table*}[htp]
	\centering
	\caption{Comparison Results of LSTM, SA, CNNMP, and SEMP}
    \renewcommand{\arraystretch}{1.1}
    \label{tab2}
	\begin{tabular}{|c|c|c|c|c|c|c|c|c|}
		\hline
		\multirow{2}*{\textbf{Method}} & \multicolumn{4}{c|}{\textbf{Average Reward [Avg / Var]}} &  \multicolumn{4}{c|}{\textbf{Collision Rate (\%) [Avg / Var]}} \\
		\cline{2-9}
		\multicolumn{1}{|c|}{} & {$n$ = 4} & {$n$ = 6} & {$n$ = 8} & {$n$ = 10} & $n$ = 4 & $n$ = 6 & $n$ = 8 & $n$ = 10 \\
		\hline
		LSTM~\cite{ref16} & -70.99/330.43 & -76.07/292.45 & -81.89/97.87 & -92.70/152.85 &
                            0.179/0.019 & 0.300/0.058 & 0.325/0.026 & 0.429/0.029 \\
		\hline
		SA~\cite{ref29} & -49.25/97.86 & -59.01/104.62 & -67.20/101.59 & -79.72/155.76 &
                            0.282/0.120 & 0.272/0.016 & 0.375/0.025 & 0.554/0.038 \\
        \hline
		CNNMP~\cite{ref26} & -48.88/127.72 & -57.51/133.02 & -63.39/112.21 & -72.98/122.63 &
                            0.215/0.048 & 0.282/0.026 & 0.294/0.025 & 0.404/0.036 \\
		\hline
         SEMP (Ours) & \textbf{-36.95/58.32} & \textbf{-45.65/79.99} & \textbf{-51.91/60.34} & \textbf{-60.54/49.67} &
         \textbf{0.140/0.011} & \textbf{0.192/0.016} & \textbf{0.285/0.021} & \textbf{0.317/0.021} \\
		\hline
	\end{tabular}

\end{table*}

The learning curves of the four above approaches are shown in
Figure~\ref{figurelabe4}. As can be seen, the LSTM scheme has the worst performance in terms of
both learning efficiency and the obtained average reward. One  reasonable
explanation is that LSTM has too many parameters to optimize. Besides, the
learning curves of SA and CNNMP are almost overlapped, indicating that the two
schemes have similar performance. Additionally, the average reward obtained by
SEMP grows as fast as CNNMP in the early stage. However, with the increase of
training episodes, the average reward of SEMP still gradually increases, and
finally obtains a higher reward. This comparison results validate that the
attention mechanism with SE blocks used by SEMP is effective.
Overall, SEMP finally obtains the highest reward with a similar learning
efficiency. This result demonstrates that our SEMP scheme has advantage over the
three benchmarks.

\begin{figure}[h!]
	\centering
	\includegraphics[width=3.3in]{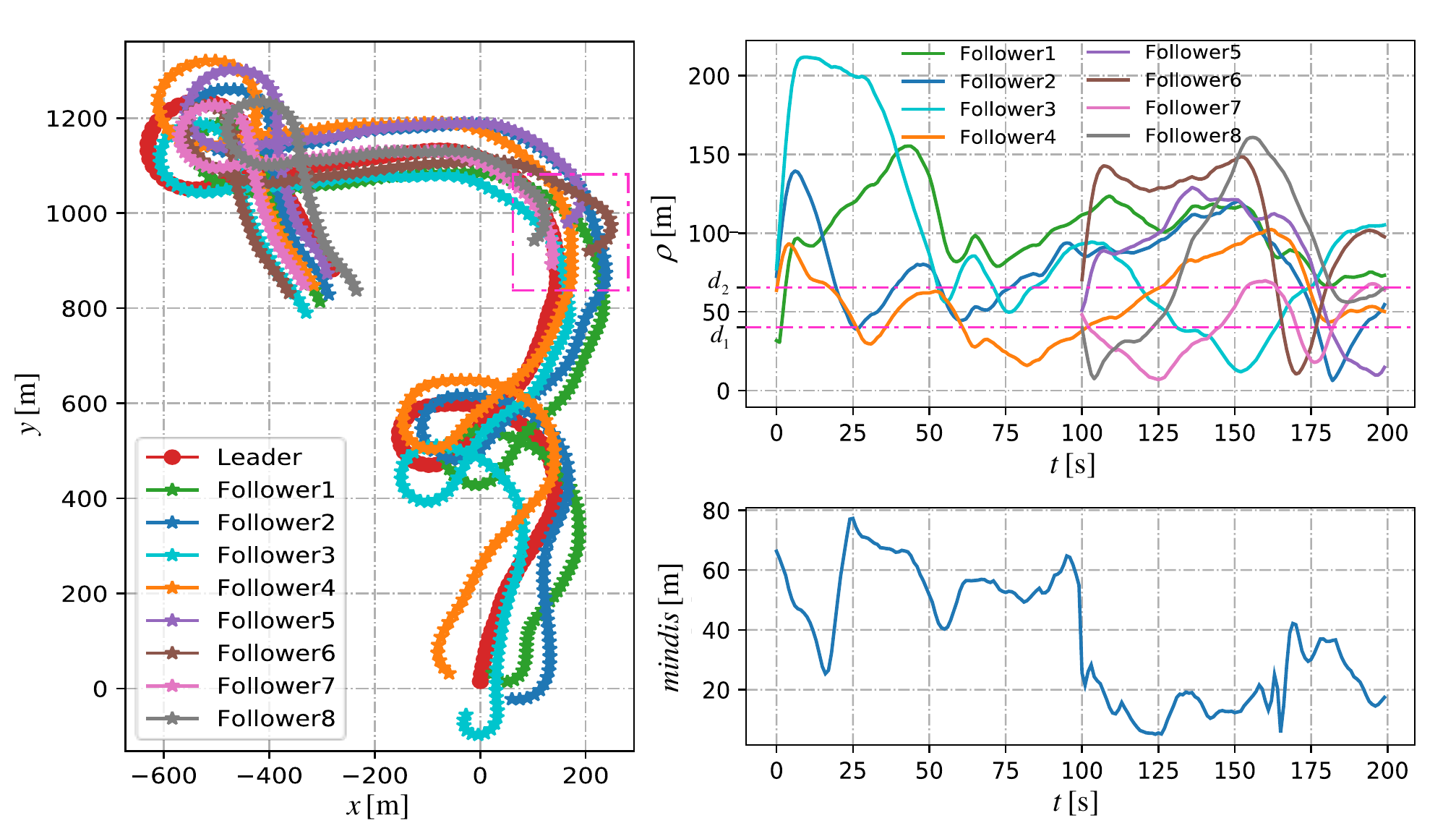}
	%\framebox{\parbox{3in}{fig1.jpg}}
	%\includegraphics[scale=1.5]{fig2.png}
	\caption{The visualized results of the eight followers following the leader using the learned SEMP policy, i.e., trajectories (\textbf{left}), the distances ($\rho$) to the leader (\textbf{top-right}), and the minimum distance ($mindis$) among the followers (\textbf{bottom-right}).}
	\label{figurelabe5}
\end{figure}

After training, we tested the control policy learned by the PS-CACER algorithm
with the SEMP scheme (hereafter referred to as the learned SEMP policy for short)
in a flocking task with eight followers lasting for 200 seconds. Initially,
the leader was followed by four followers (Follower \#1 $\sim$ \#4 in Figure~\ref{figurelabe5}).
Since the time step 100, the other four followers (Follower \#5 $\sim$ \#8 in Figure~\ref{figurelabe5})
joined the flock simultaneously. The testing results are visualized in Figure~\ref{figurelabe5}.
The generated trajectories illustrate that the eight followers can successfully keep up with the leader while avoiding collisions with each other. This result indicates that the learned SEMP policy can handle a variable number of followers.

\begin{figure}[t!]
  \centering
  \includegraphics[width=3.3in]{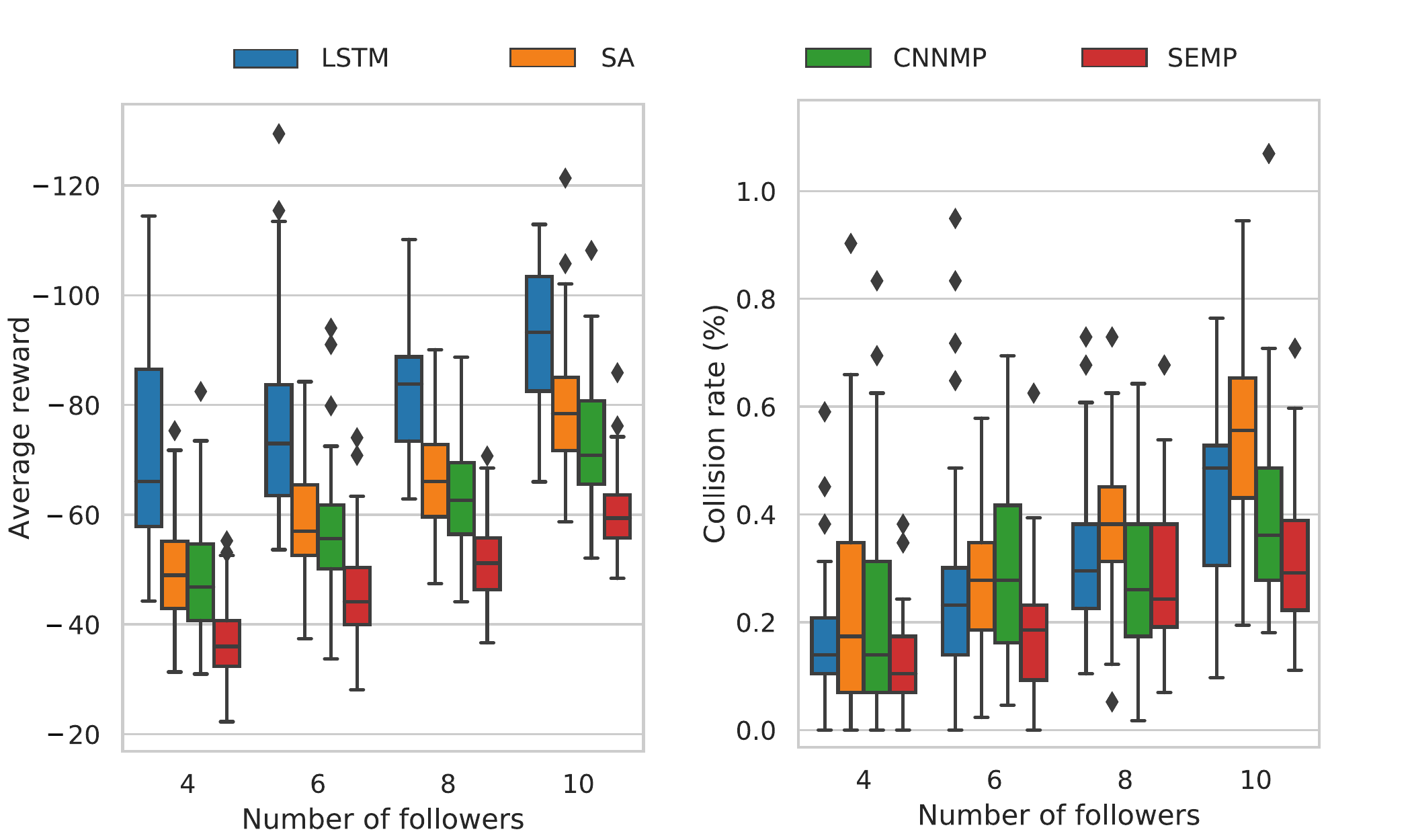}
      %\framebox{\parbox{3in}{fig1.jpg}}
      %\includegraphics[scale=1.5]{fig2.png}
  \caption{Box-and-whisker plot for comparison results of LSTM, SA, CNNMP, and SEMP.
\textbf{Left}: Average reward. \textbf{Right}: Collision rate.}
  \label{figurelabe6}
\end{figure}

To further quantitatively evaluate the performance of the learned SEMP policy,
we defined \textit{collision rate} as a metric aside from the average reward.
We considered that a collision would happen if the distance between two followers was
less than a certain value, e.g., 2 meters. Thus, the \textit{collision rate} meant the percentage of
being too close (i.e., less than 2 meters) among the followers during the testing
episode. We compared the learned SEMP policy with three baselines by averaging the
results of 200 episode. The parameter settings are the same as the training phase, but the maximum time step
($N_{s}$) for each episode was set to 180. Table II and Figure~\ref{figurelabe6} compares
the results of the four methods.

As can be seen from Table~\ref{tab2}, the learned policies with one set of network parameters can adapt to a
dynamic squad of UAVs, without the need of retraining in a new environment with
different number of followers. When the number of followers increases, the
average reward decreases and the collision rate increases. The reason is that
increasing the number of followers leads to a higher probability of collision.
However, compared with the other three baselines, our SEMP method enables the
follower to obtain the largest average reward, the lowest collision rate, and the
lowest variance, regardless of the number of followers. This result shows that
the SEMP method outperforms the three existing state-of-the-art methods.

\subsection{Semi-Physical Simulation}

In addition to the numerical simulation, we also tested the generalization
performance of the learned control policy by conducting a hardware-in-loop (HIL)
experiment in a high fidelity semi-physical simulation system
\cite{ref14},~\cite{ref15}. Instead of the kinematic model of fixed-wing UAVs
with stochastic disturbances used in the training phase, we selected the
professional tool X-Plane
10\footnote{https://www.x-plane.com/manuals/desktop/} as the flight simulator for the testing.
 X-Plane 10 can simulate complex environmental conditions,
such as weather changes and wind disturbances.

In this experiment, the five followers used the learned SEMP policy to flock
with the leader lasting for 200 time steps (seconds). At each time step, the
leader selected its roll-angle setpoint according to the pre-planned paths and
changed its velocity randomly, then broadcasted its state information to all the followers.
Each follower used the actor network with the trained parameters to
determine its steering commands according to its own state, the leader's state,
and the other followers' states.

\begin{figure}[t!]
  \centering
  \includegraphics[width=3.0in]{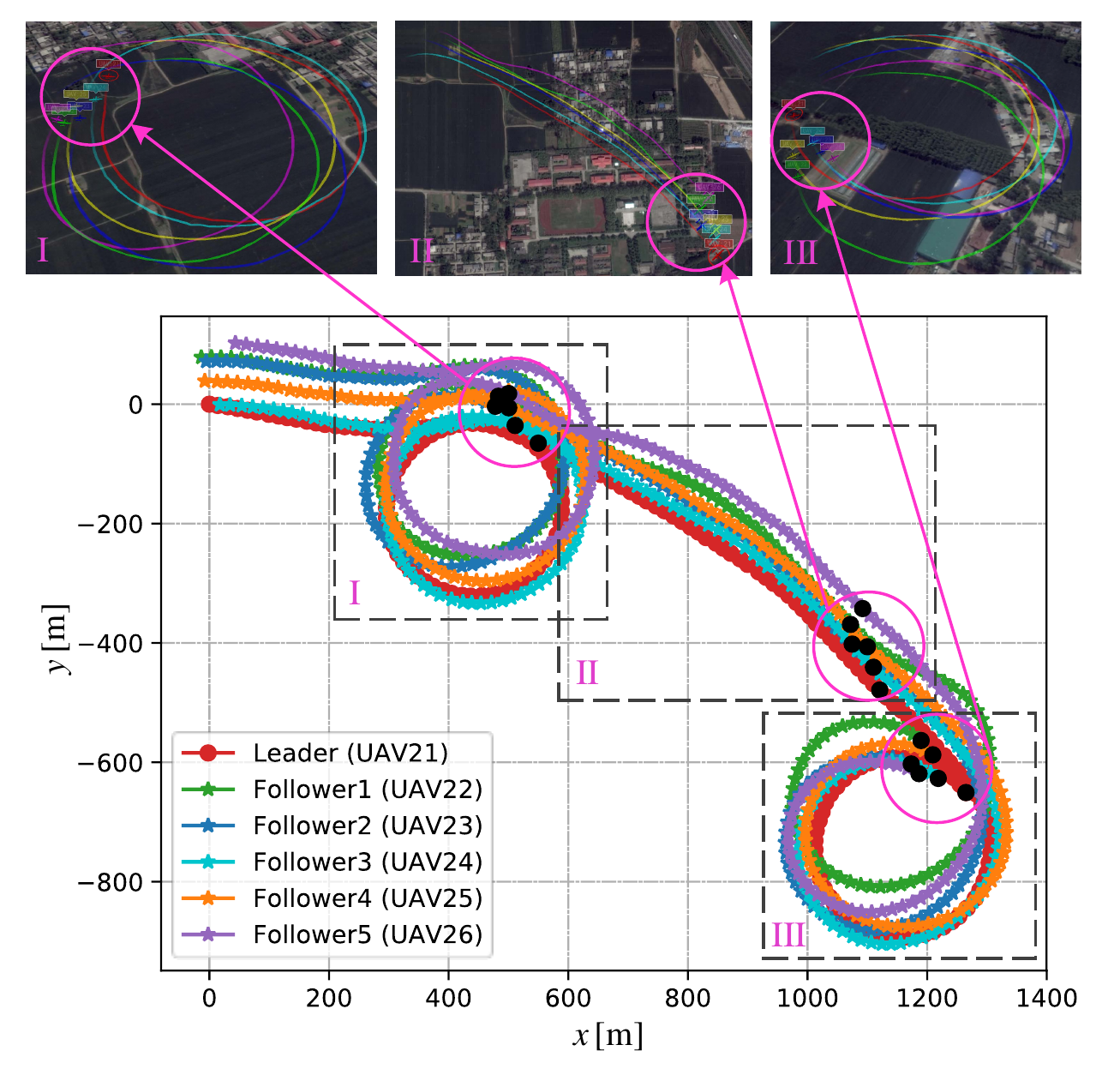}
      %\framebox{\parbox{3in}{fig1.jpg}}
      %\includegraphics[scale=1.5]{fig2.png}
  \caption{The trajectory results in the semi-physical simulation. Note that the three
representative snapshots displayed on the top are captured from the ground
control station. The top-left and top-right snapshots show a third person
perspective of the UAV flock, while the top-middle snapshot is top view.}
  \label{figurelabe7}
\end{figure}

\begin{figure}[t!]
  \centering
  \includegraphics[width=3.0in]{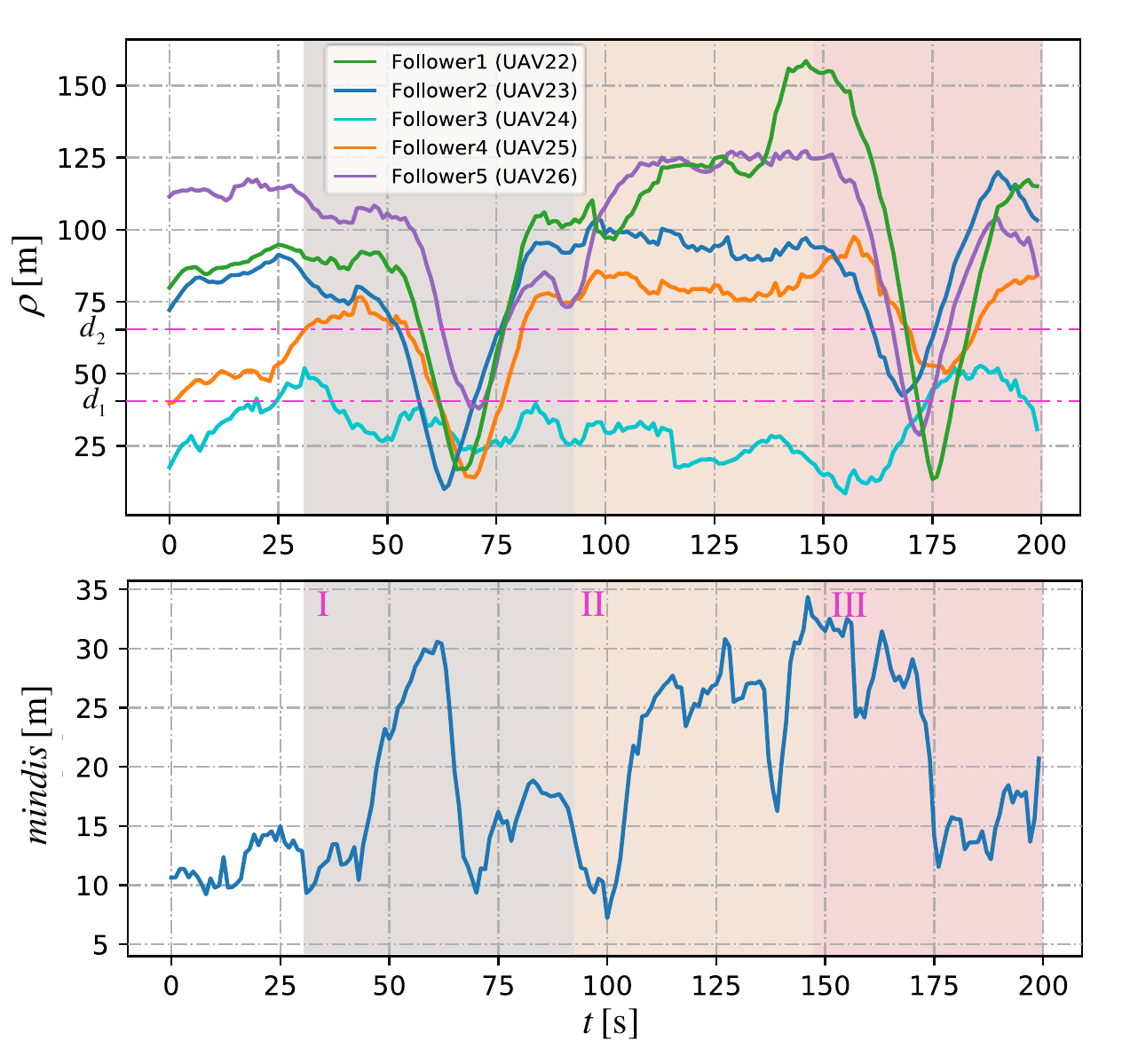}
      %\framebox{\parbox{3in}{fig1.jpg}}
      %\includegraphics[scale=1.5]{fig2.png}
  \caption{The distances between the leader and five followers ($\rho$) and the
minimum distance among five followers ($mindis$) in the semi-physical
simulation.}
  \label{figurelabe8}
\end{figure}

The trajectory results are depicted in Figure~\ref{figurelabe7}; the distances between the
leader and the five followers ($\rho$) as well as the minimum distance among
the five followers ($mindis$) are shown in Figure~\ref{figurelabe8}. As can be seen, the
distances between the leader and the followers were maintained around 75$\,\rm{m}$
most of the time (the average distance is 74.44$\,\rm{m}$). This means that the five followers were able to keep up with
the leader steadily, even if the leader changed its heading sharply. More
importantly, the minimum value of $mindis$ is larger than 2$\,\rm{m}$, which means
that there were no collisions among the followers during this experiment. The
above results demonstrate that the proposed SEMP method enables the followers to
avoid collisions between each other while flocking with the leader. We note that the control policy employed by the five followers in this
semi-physical simulation is the SEMP policy learned from the previous numerical
simulation, without any parameter finetuning. This result demonstrates that the
learned SEMP policy can be directly generalized to new situations.

\section{Conclusion}
\label{sec5}

In this paper, we have proposed a deep reinforcement learning algorithm to
solve the leader-follower flocking control and collision avoidance problem for a dynamic
squad of fixed-wing UAVs. Specifically, we have designed a decentralized
DRL-based framework where the collision avoidance policy is integrated into the
flocking controller for a variable number of UAVs. Then, we have proposed the
PS-CACER algorithm for training multiple agents, in which a customized embedding
module, SEMP, is integrated to handle the variable-length input for deep neural
networks. With this module, the learned policies are adaptive to the changes of the number of UAVs without the
need of retraining. Numerical simulation results have demonstrated that the
proposed method outperforms three existing state-of-the-art methods, i.e.,
LSTM, SA, and CNNMP. Finally, the learned policy can be directly transferred to a
semi-physical simulation without any parameter finetuning. In the future, we will
further evaluate our method with fixed-wing UAVs in real-world environments.

\addtolength{\textheight}{-12cm}   % This command serves to balance the column lengths

%%%%%%%%%%%%%%%%%%%%%%%%%%%%%%%%%%%%%%%%%%%%%%%%%%%%%%%%%%%%%%%%%%%%%%%%%%%%%%%%

%%%%%%%%%%%%%%%%%%%%%%%%%%%%%%%%%%%%%%%%%%%%%%%%%%%%%%%%%%%%%%%%%%%%%%%%%%%%%%%%

%%%%%%%%%%%%%%%%%%%%%%%%%%%%%%%%%%%%%%%%%%%%%%%%%%%%%%%%%%%%%%%%%%%%%%%%%%%%%%%%

\end{document}